\documentclass{elsart}
\usepackage{graphics,natbib,graphicx,psfig,amssymb} 
\journal{New Astronomy}
\input psfig.sty
\begin{document}
\begin{frontmatter}
\title{A new absolute magnitude calibration with {\it 2MASS} for cataclysmic variables}
\author[istanbul]{T. Ak\corauthref{cor}},
\corauth[cor]{corresponding author.}
\ead{tanselak@istanbul.edu.tr}
\author[istanbul]{S. Bilir},
\author[istanbul]{S. Ak},
\author[israel]{and A. Retter}
\address[istanbul]{Istanbul University, Faculty of Sciences, Department 
of Astronomy and Space Sciences, 34119 University, Istanbul, Turkey}
\address[israel]{P.O. Box 4264, Shoham, 60850, Israel \\}

\begin{abstract}
Using reliable trigonometric measurements, we find that the absolute magnitude 
of cataclysmic variables depends on the orbital period and de-reddened $(J-H)_{0}$ and $(H-K_{s})_{0}$ 
colours of {\em 2MASS} (Two Micron All Sky Survey) photometric system. The calibration equation covers 
the ranges $0.032^{d} < P_{orb} \leq 0.454^{d}$, $-0.08 < (J-H)_{0} \leq 1.54$, 
$-0.03 < (H-K_{s})_{0} \leq 0.56$ and $2.0 < M_{J} < 11.7$; It is based on trigonometric parallaxes 
with relative errors of $(\sigma_{\pi}/\pi) \leq 0.4$. By using the period-luminosity-colours (PLCs) 
relation, we estimated the distances of cataclysmic variables with orbital periods and {\em 2MASS} 
observations and compared them with distances found from other methods. We suggest that the PLCs relation 
can be a useful statistical tool to estimate the distances of cataclysmic variables.
\end{abstract}

\end{frontmatter}

\section{Introduction}
Cataclysmic variables (hereafter referred to as CVs) are short period interacting binary 
stars in which a red-dwarf, the secondary star, overflows its Roche lobe and transfers 
matter to a white dwarf typically via an accretion disc. A bright spot is formed in the 
location where the matter stream impacts the accretion disc. In CVs that have strongly 
magnetic white dwarf pimaries the accreting matter can not construct an accretion 
disc, instead, the accretion is maintained through accretion columns above the magnetic 
poles of the white dwarf. For a detailed description of the CV phenomenon and its subclasses 
see \citet{Warner1995} and \citet{Hellier2001}.

Although distances of CVs are needed to improve and constrain physical models, reliable distance 
measurements can be only found for a few systems. Different methods have been 
used to measure CV distances (\citealt{Thorstensen2003}). However, all but trigonometric parallaxes 
yield rough distance estimates. The most promising method for determining CV distances 
has been to make use of the properties of the secondary star (\citealt{Bailey1981}; Sproats, 
Howell $\&$ Mason \citeyear{Sproatsetal1996}). This method assumes that all $K$-band 
emission originates from the secondary star. Almost all secondary stars in CVs lie on 
or near the ZAMS (Zero Age Main Sequence) for near-solar metallicity within the uncertainties 
\citep{Warner1995,Beuermannetal1998,Beuermann2000,KolbandBaraffe2000}. However, if one tries 
to measure the absolute magnitude of the secondary star, contributions to the light from the 
other components contaminate the results. Therefore, only a lower limit to the distance can be 
obtained by using this method due to the effects of the disc and the irradiated area of the 
secondary star (Berriman, Szkody $\&$ Capps \citeyear{Berrimanetal1985}). Almost all distance 
estimation methods attempts to use properties of a component of the system such as surface 
brightness of the secondary star (\citealt{Bailey1981}), spectra of the white dwarf 
(\citealt{Sionetal1995}; \citealt{UrbanandSion2006}) and $M_{V}-P_{orb}$ relationship of dwarf 
novae at outburst (\citealt{Warner1995}; \citealt{Harrisoneta2004}). However, none of them can 
give a distance as precise as trigonometric parallax due to the contaminations from the other 
components and the lack of the information about the individual contributions of the components 
to the observed light. 

It is accepted that the most reliable distances are obtained from trigonometric parallaxes. 
However, this trigonometric parallax method can only be applied to the closest objects due to 
observational constraints. First precise trigonometric parallax measurements of the brightest 
CVs came from Hipparcos Satellite (\citealt{Duerbeck1999}). Trigonometric parallaxes of some 
CVs were measured by using Hubble Space Telecope's Fine Guidence Sensor 
(\citealt{McArthuretal1999,McArthuretal2001}; \citealt{Beuermannetal2003a,Beuermannetal2003b}; 
\citealt{Harrisoneta2004}), as well. In addition, Thorstensen (\citeyear{Thorstensen2003}) 
measured trigonometric parallaxes of 14 CVs from ground-based observations.

A relationship of absolute magnitude in any wavelength interval with the orbital period and 
at least one colour of the system can be a very useful tool to estimate distances of binary stars. 
This method was applied for W UMa-type binary stars 
(\citealt{RucinskiandDuerbeck1997,Rucinski2004}). If there is such a relationship for CVs as well, 
their distances can be statistically estimated to a certain precision. For this, 
{\em 2MASS} magnitudes and colours can be a good choice since most of the light in these 
photometric bands comes from the secondary star and the effect of the interstellar reddening 
for $J$, $H$ and $K$ bands is weaker than that in visual wavelengths. Although we can not measure 
the absolute magnitude of the secondary star to a good precision, at least, we know that the 
secondary star in a CV can be considered as the least-active component of the system. 

We should state that our aim is $\it not$ to measure the absolute magnitudes of the secondary 
stars in CVs. In this study, we first estimate the systemic $J$-band absolute magnitudes $M_{J}$ 
of the closest CVs using reliable trigonometric parallaxes by assuming that the light in $J$, 
$H$ and $K$ bands comes from the system as a $\it whole$. Then, we find the dependence of the
absolute magnitude on the orbital period $P_{orb}$ and de-reddened colours $(J-H)_{0}$ and 
$(H-K)_{0}$ to derive an absolute magnitude calibration for CVs with {\em 2MASS} photometric 
system.

\section{The Data}

Our data sample consists of CVs with trigonometric parallax ($\pi$) errors smaller than 
($\sigma_{\pi}/\pi$)$\leq$0.4. The 27 systems listed in Table 1 include CVs with orbital 
periods shorter than $\sim$12 hr since a CV with orbital period longer than this limit possibly 
contains a secondary star on its way to becoming a red giant (\citealt{Hellier2001}). Dwarf novae 
and nova-like systems were selected from \citet{Duerbeck1999}, \citet{McArthuretal1999,McArthuretal2001}, 
\citet{Beuermannetal2003a,Beuermannetal2003b}, \citet{Thorstensen2003} and \citet{Harrisoneta2004}. 
\citet{Duerbeck1999} obtained trigonometric parallaxes of four novae however, we included only two 
of them in our sample. The orbital period of T CrB ($P_{orb}\sim$228 days), which is classified 
as a recurrent nova in \citet{Downesetal2001} catalogue, is longer than our upper limit. In addition, 
a comparison of Tables 1 and 2 in \citet{Duerbeck1999} shows that the distance of HR Del evaluated 
from the shell expansion parallax method is very different from its distance found from trigonometric 
parallax. Thus, we excluded these two systems from our sample. Table 1 lists the systems used in the 
analysis. Our data sample consists of 14 dwarf novae, 11 nova-like stars and two novae. These are CVs 
with the most precise distance estimates ever found in the literature. 

$J$, $H$ and $K_{s}$ magnitudes were taken from the Point-Source Catalogue and Atlas 
(\citealt{Cutrietal2003}; \citealt{Skrutskieetal2006}) which is based on the {\em 2MASS} (Two Micron All 
Sky Survey) observations. The {\em 2MASS} photometric system comprises Johnson's $J$ (1.25 $\mu$m) 
and $H$ (1.65 $\mu$m) bands with the addition of $K_{s}$ (2.17 $\mu$m) band, which is bluer than Johnson's 
$K$-band. Although we study the most closest CVs ever known, the total interstellar absorption in the 
direction of the star should be taken into account. We used the equations of 
\citet{FiorucciandMunari2003} for the determination of the total absorption for $J$, $H$ and 
$K_{s}$ bands, i.e. $A_{J}=0.887\times E(B-V)$, $A_{H}=0.565\times E(B-V)$ and 
$A_{K_{s}}=0.382\times E(B-V)$, respectively. 

Fortunately, the $E(B-V)$ colour excesses of many CVs were estimated by \citet{BruchandEngel1994} and 
\citet{Harrisoneta2004}. Bruch $\&$ Engel's catalogue includes the systems whose $E(B-V)$ values 
were given in \citet{Verbunt1987} and \citet{LaDous1991}. Our primary $E(B-V)$ source is 
\citet{Harrisoneta2004}. If we did not find the $E(B-V)$ value of a star in their study, we returned 
to \citet{BruchandEngel1994}. Unfortunately, colour excesses of GP Com, GW Lib, EF Eri and V893 Sco 
were not given in the sources mentioned above. Thus, we calculated their colour excesses from 
Schlegel, Finkbeiner $\&$ Davis (\citeyear{Schlegeletal1998}) maps by using NASA Extragalactic 
Database \footnote{http://nedwww.ipac.caltech.edu/forms/calculator.html}. Since these are relatively 
close systems, the colour excesses found from \citet{Schlegeletal1998} need to be reduced according 
to the stellar distance. In order to do this, we used the $E_{\infty}(B-V)$ colour excess in the 
Galactic latitude ($b$) and longitude ($l$) for the model from \citet{Schlegeletal1998}. The total 
absorption for the model was evaluated from

\begin{equation}
A_{\infty}(b)=3.1E_{\infty}(B-V).
\end{equation}
The total absorption for the distance $d$ to the star is calculated as following
\citep{BahcallandSoneira1980}

\begin{equation}
A_{d}(b)=A_{\infty}(b)\Biggl[1-exp\Biggl(\frac{-\mid d~sin(b)\mid}{H}\Biggr)\Biggr],
\end{equation}
where $H$ is the scaleheight for the interstellar dust which is adopted as 100 pc. 
Finally, the colour excess for a star at the distance $d$ is estimated from

\begin{equation}
E_{d}(B-V)=A_{d}(b)~/~3.1.
\end{equation}

Once we obtained the apparent magnitudes ($J$, $H$ and $K_{s}$), total absorption ($A_{J}$, 
$A_{H}$ and $A_{K_{s}}$) and distance $d=1/\pi$ for a CV, the absolute magnitudes ($M_{J}$, 
$M_{H}$ and $M_{K_{s}}$) of the system were easily calculated from the well known 
distance-modulus formula, i.e. $M_{J}=J-5\log d+5-A_{J}$. The calculated $M_{J}$ 
values are listed in Table 1.


\begin{table*}
\scriptsize{
\begin{center}
\caption{\scriptsize{The data sample. Types and orbital periods ($P_{orb}$) were taken from 
\citet{Downesetal2001}. DN, NL and N denote dwarf nova, nova-like star and nova, respectively. 
$J$, $H$ and $K_{s}$ magnitudes were collected from the {\em 2MASS} Point Source Catalogue  
\citep{Cutrietal2003}. $\pi$ denotes parallax, $E(B-V)$ colour excess and 
$M_{J}$ absolute magnitude in $J$-band.}}
\begin{tabular}{lccccccccc}
\hline
Name      & Type & $P_{orb}$ & $J$ & $J-H$ & $H-K_{s}$ & $\pi$ &  $\sigma_{\pi}/\pi$ & $E(B-V)$ & $M_{J}$  \\
          &      & (days)       &   &     &     & (mas) &                     &          &         \\
\hline
GP Com    &NL&0.0323&15.72$\pm$0.07 & 0.10$\pm$0.15 & 0.48$\pm$0.21 & 14.8$^{f}$  & 0.09 & 0.01$^{i}$ & 11.55$\pm$0.26  \\
GW Lib    &DN&0.0533&16.19$\pm$0.09 & 0.60$\pm$0.15 & 0.19$\pm$0.22 & 11.5$^{f}$  & 0.18 & 0.06$^{i}$ & 11.44$\pm$0.48  \\
EF Eri    &NL&0.0563&17.21$\pm$0.24 & 1.54$\pm$0.28 & 0.30$\pm$0.22 &  5.5$^{f}$  & 0.36 & 0.01$^{i}$ & 10.90$\pm$1.03  \\
WZ Sge    &DN&0.0567&14.86$\pm$0.04 & 0.30$\pm$0.06 & 0.56$\pm$0.08 & 22.97$^{a}$ & 0.01 & 0.00$^{a}$ & 11.67$\pm$0.06  \\
T Leo     &DN&0.0588&14.77$\pm$0.04 & 0.44$\pm$0.07 & 0.51$\pm$0.08 & 10.2$^{f}$  & 0.11 & 0.00$^{h}$ &  9.82$\pm$0.28  \\
VY Aqr    &DN&0.0631&15.28$\pm$0.05 & 0.42$\pm$0.11 & 0.27$\pm$0.13 & 11.2$^{f}$  & 0.12 & 0.07$^{h}$ & 10.47$\pm$0.31  \\
EX Hya    &DN&0.0682&12.27$\pm$0.02 & 0.32$\pm$0.04 & 0.26$\pm$0.04 & 15.50$^{b}$ & 0.02 & 0.00$^{a}$ &  8.23$\pm$0.07  \\
V893 Sco  &DN&0.0760&13.22$\pm$0.03 & 0.31$\pm$0.04 & 0.23$\pm$0.04 &  7.4$^{f}$  & 0.27 & 0.12$^{i}$ &  7.46$\pm$0.61  \\
SU UMa    &DN&0.0764&11.78$\pm$0.02 & 0.05$\pm$0.03 & 0.06$\pm$0.03 &  7.4$^{f}$  & 0.20 & 0.00$^{h}$ &  6.12$\pm$0.46  \\
YZ Cnc    &DN&0.0868&13.17$\pm$0.02 & 0.22$\pm$0.03 & 0.12$\pm$0.03 &  3.34$^{a}$ & 0.10 & 0.00$^{a}$ &  5.78$\pm$0.24  \\
AM Her    &NL&0.1289&11.70$\pm$0.02 & 0.51$\pm$0.03 & 0.19$\pm$0.03 & 13.0$^{f}$  & 0.09 & 0.00$^{j}$ &  7.27$\pm$0.22  \\
V603 Aql  &N &0.1385&11.70$\pm$0.03 & 0.19$\pm$0.04 & 0.16$\pm$0.05 &  4.21$^{g}$ & 0.38 & 0.08$^{h}$ &  4.76$\pm$0.85  \\
V1223 Sgr &NL&0.1402&12.81$\pm$0.02 & 0.07$\pm$0.04 & 0.10$\pm$0.04 &  1.95$^{e}$ & 0.08 & {\bf 0.15$^{a}$} &  {\bf 4.12$\pm$0.20}  \\
RR Pic    &N &0.1450&12.46$\pm$0.02 & 0.06$\pm$0.02 & 0.14$\pm$0.03 &  2.46$^{g}$ & 0.32 & 0.02$^{h}$ &  4.39$\pm$0.72  \\
U Gem     &DN&0.1769&11.65$\pm$0.02 & 0.58$\pm$0.03 & 0.24$\pm$0.03 &  9.96$^{a}$ & 0.03 & 0.00$^{a}$ &  6.64$\pm$0.09  \\
SS Aur    &DN&0.1828&12.70$\pm$0.02 & 0.43$\pm$0.03 & 0.27$\pm$0.03 &  5.99$^{a}$ & 0.05 & {\bf 0.03$^{a}$} &  {\bf 6.56$\pm$0.13}  \\
IX Vel    &NL&0.1939& 9.12$\pm$0.03 & 0.14$\pm$0.04 & 0.15$\pm$0.03 & 10.38$^{g}$ & 0.09 & 0.01$^{h}$ &  4.20$\pm$0.23  \\
V3885 Sgr &NL&0.2071& 9.96$\pm$0.03 & 0.22$\pm$0.04 & 0.12$\pm$0.04 &  9.11$^{g}$ & 0.19 & 0.02$^{h}$ &  4.74$\pm$0.44  \\
TV Col    &NL&0.2286&13.20$\pm$0.03 & 0.37$\pm$0.03 & 0.14$\pm$0.04 &  2.70$^{d}$ & 0.04 & {\bf 0.05$^{a}$} &  {\bf 5.31$\pm$0.11}  \\
RW Tri    &NL&0.2319&11.94$\pm$0.02 & 0.36$\pm$0.03 & 0.12$\pm$0.03 &  2.93$^{c}$ & 0.09 & {\bf 0.26$^{a}$} &  {\bf 4.04$\pm$0.22}  \\
RW Sex    &NL&0.2451&10.32$\pm$0.03 & 0.18$\pm$0.04 & 0.08$\pm$0.03 &  3.46$^{g}$ & 0.29 & 0.02$^{h}$ &  3.00$\pm$0.66  \\
AH Her    &DN&0.2581&11.81$\pm$0.02 & 0.33$\pm$0.03 & 0.10$\pm$0.03 &  3$^{f}$    & 0.40 & 0.03$^{h}$ &  4.17$\pm$0.89  \\
SS Cyg    &DN&0.2751& 8.52$\pm$0.02 & 0.16$\pm$0.03 & 0.06$\pm$0.03 &  6.06$^{a}$ & 0.06 & {\bf 0.04$^{a}$} &  {\bf 2.39$\pm$0.15}  \\
Z Cam     &DN&0.2898&11.57$\pm$0.03 & 0.53$\pm$0.03 & 0.19$\pm$0.04 &  8.9$^{f}$  & 0.17 & 0.02$^{h}$ &  6.31$\pm$0.40  \\
RU Peg    &DN&0.3746&11.07$\pm$0.02 & 0.44$\pm$0.03 & 0.16$\pm$0.03 &  3.55$^{a}$ & 0.06 & 0.00$^{a}$ &  3.82$\pm$0.15  \\
AE Aqr    &NL&0.4117& 9.46$\pm$0.02 & 0.54$\pm$0.03 & 0.14$\pm$0.03 &  9.8$^{g}$  & 0.24 & 0.00$^{h}$ &  4.42$\pm$0.54  \\
QU Car    &NL&0.4540&10.97$\pm$0.02 & 0.12$\pm$0.03 & 0.14$\pm$0.03 &  1.64$^{g}$ & 0.32 & 0.11$^{h}$ &  2.04$\pm$0.72  \\
\hline
\end{tabular}
\end{center}
{a: \citet{Harrisoneta2004}, b: \citet{Beuermannetal2003a}, c: \citet{McArthuretal1999}, 
d: \citet{McArthuretal2001}, e: \citet{Beuermannetal2003b}, f: \citet{Thorstensen2003}, 
g: \citet{Duerbeck1999}, h: \citet{BruchandEngel1994}, i: \citet{Schlegeletal1998}, 
j: \citet{LaDous1991}}
}
\end{table*}

\section{Analysis}

We used the data in Table 1 to derive an absolute magnitude calibration for CVs with {\em 2MASS}. 
In order to find the dependence of the absolute magnitude $M_{J}$ on the orbital period $P_{orb}$, 
and colours $(J-H)_{0}$ and $(H-K_{s})_{0}$, we used a fit equation in the following form

\begin{displaymath}
M_{J}=a+b~\log P_{orb}(day)+c~(J-H)_{0}+d~(H-K)_{0},
\end{displaymath}
whose least square coefficients and their 1-$\sigma$ errors are given in Table 2. The subscript 
'0' indicates de-reddened magnitudes. The $M_{J}$ calibration that utilizes de-reddened colour 
indices $(J-H)_{0}$ and $(H-K_{s})_{0}$ and the periods can be used to predict individual values 
within an error of about $\pm$ 0.22 mag. The correlation coefficient and standard deviation of the 
calibration were estimated as $R = 0.96$ and $s = 0.88$, respectively. It should be emphasized that 
the calibration equation covers the ranges $0.032^{d} < P_{orb} \leq 0.454^{d}$, 
$-0.08 < (J-H)_{0} \leq 1.54$, $-0.03 < (H-K_{s})_{0} \leq 0.56$ and $2.0 < M_{J} < 11.7$; It is 
based on trigonometric parallaxes with relative errors $(\sigma_{\pi}/\pi) \leq 0.4$. A comparison 
of the fit values of absolute magnitudes, $M_{Jc}$, calculated from this period-luminosity-colours 
(PLCs) relation with the observed $M_{J}$ absolute magnitudes is shown in Figure 1.


\begin{figure}
\begin{center}
\includegraphics[scale=0.375, angle=0]{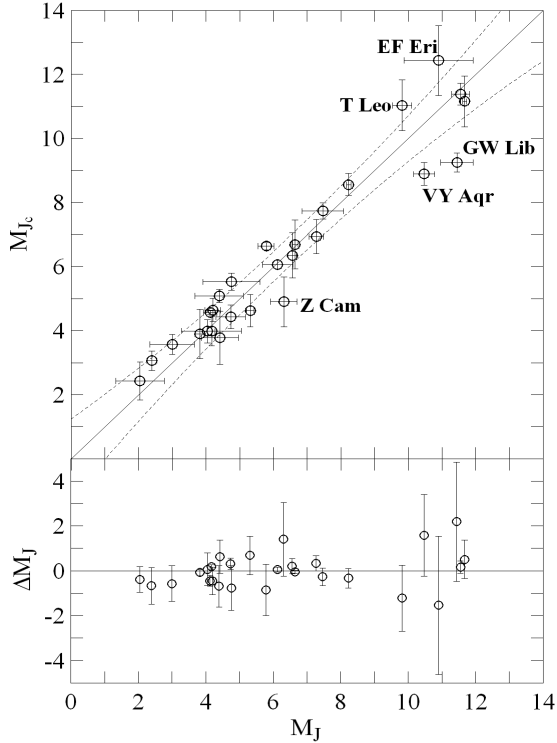}
\caption[] {\small A comparison of the absolute magnitudes ($M_{Jc}$) calculated 
from the PLCs relation with the estimated $M_{J}$ absolute magnitudes in Table 1. Upper and lower 
confidence limits of 99$\%$ are shown with dotted lines. The bottom panel displays the residuals 
from the fit. The diagonal line represents the equal values. Largely scattered systems are shown.}
\end{center}
\end{figure}


\begin{table}
\begin{center}
\caption{Coefficients of the calibration equation.}
\small
\begin{tabular}{ccccc}
\hline
Coefficient &     $a$     &       $b$    &    $c$      &    $d$      \\
\hline
            &   {\bf ~-0.894}   &   {\bf ~-5.721}    &   {\bf ~~2.598}    &  {\bf ~~7.380}     \\
$\sigma$    & {\bf $\pm$0.522}  &  {\bf $\pm$0.705}  &  {\bf$\pm$0.610}   &  {\bf $\pm$1.711}   \\
\hline
\end{tabular}
\end{center}
\end{table}


\begin{figure}
\begin{center}
\includegraphics[scale=0.35, angle=0]{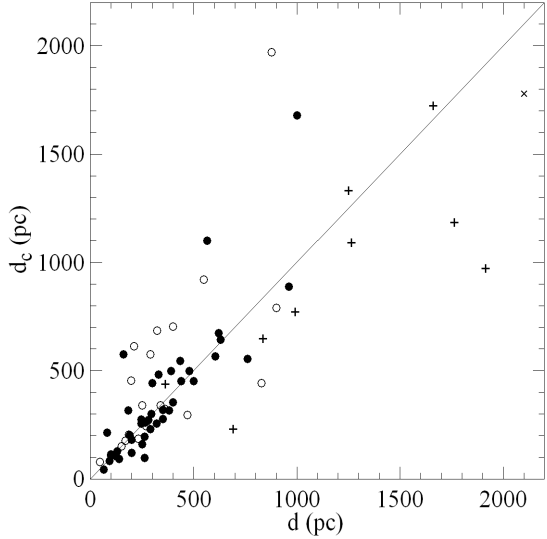}
\caption[] {\small A comparison of the distances ($d_{c}$) calculated using the absolute 
magnitudes ($M_{Jc}$) from the PLCs relation, with these distances ($d$) estimated from 
other methods. The diagonal line represents the equal values. The symbols 
$\bullet$, $\circ$ and $+$ denote dwarf novae, nova-like systems and novae respectively.
The type of the cataclysmic variable J0813+4528 is unknown and it is shown by 
$\times$ in the figure.}
\end{center}
\end{figure}

\subsection{Application to other cataclysmic variables}
We collected the list of cataclysmic variables with $E(B-V)$ values from \citet{BruchandEngel1994}. Then, we 
selected systems with orbital periods shorter than 12 hr from Downes et al.'s 
catalogue\footnote{http://archive.stsci.edu/prepds/cvcat/} (\citealt{Downesetal2001}) and with 
{\em 2MASS} observations. Systems in the application limits of the PLCs relation (see, Section 3) are 
listed in Table 3. We also collected their distances, when available, from the literature. 
For example, \citet{UrbanandSion2006} give distances of many dwarf novae estimated from 
the $M_{v}-P_{orb}$ relation. Most of the distances mentioned in Table 3 were taken 
from \citet{UrbanandSion2006}. $E(B-V)$ values and distances of EZ Del, LL Lyr, RY Ser, CH UMa 
and SDSS J081327.07+452833.0 were taken from \cite{Thorstensenetal2004}, as well.

Distances of several systems were calculated by \citet{Sproatsetal1996} using Bailey's method (\citealt{Bailey1981}). 
However, their $E(B-V)$ values were not listed in \citet{BruchandEngel1994},
{\bf except for UV Per and AR And.} Among them, we selected the 
systems located above the Galactic latitude $\mid b \mid$ $\geq$ $20^{o}$ and assumed $E(B-V)=0$ for them. 
This is a reasonable assumption since in these latitudes the interstellar absorption is very small.
We have taken the distances from \citet{Sproatsetal1996} calculated under the assumption that 
the percentage of the $K$-band light contributed by the secondary star is {\bf 50}, i.e. {\bf $K\%$=50.}

{\bf Distances of V1500 Cyg and V533 Her in Table 3 were determined by the shell parallax 
method (\citealt{Cohen1985}). For V1500 Cyg, we calculated the average distance of the nova collected 
from different sources. \citet{Selvelli2004} calculated the $E(B-V)$ colour excesses and distances 
of some old novae from the UV spectra. We included these novae with orbital periods and {\em 2MASS} 
observations in Table 3.} Finally, \citet{Patterson1984} listed distances of many CVs based on various 
methods, such as detection of the secondary, proper motion, $M_{v}-EW(H_{\beta})$ relation, interstellar 
absorption, position in the Galaxy, etc. Note that none of the distances in Table 3 were determined by 
the trigonometric parallax method.  

The absolute magnitudes ($M_{Jc}$) for the systems listed in Table 3 were calculated by using 
the calibration equation given above. The distances ($d_{c}$) were evaluated from the distance modulus 
formula with absolute magnitude $M_{Jc}$ and interstellar absorption $A_{J}$. Figure 2 compares the 
distances obtained from the the PLCs relation with those collected from the literature.


\begin{table*}
\begin{center}
\tiny{
\caption{Absolute magnitudes $M_{Jc}$ and distances $d_{c}$ calculated from the PLCs relation found in this 
study. Columns 1--7 are as in Table 1. The letter $d$ denotes distances collected from the literature. The 
type of the cataclysmic variable J0813+4528 is unknown.}
\begin{tabular}{lccccccccc}
\hline
Name      &Type& $P_{orb}(d)$ &  $J$  & $J-H$ & $H-K_{s}$ &  $E(B-V)$  & $M_{Jc}$ &    $d(pc)$    &  $d_{c}(pc)$     \\
   (1)    &(2) &   (3)     &  (4)  &   (5) &    (6)    &    (7)     &     (8)     &    (9)    &    (10)      \\
\hline
DI UMa	  &DN  & 0.0546    & 15.53 &  0.27 &    0.14   &   0$^{a}$  &   8.04     & 184$^{b}$    &  316$^{+5}_{-4}$   \\
AL Com	  &DN  & 0.0567    & 16.51 &  0.33 &    0.32   &   0$^{a}$  &   9.43     & 264$^{b}$    &  261$^{+43}_{-51}$   \\
SW UMa	  &DN  & 0.0568    & 15.62 &  0.29 &    0.52   &   0$^{c}$  &  10.81     & 140$^{d}$    &   92$^{+26}_{-35}$   \\
CP Pup	  & N  & 0.0614    & 14.34 &  0.10 &    0.21   & 0.25$^{c}$ &   7.32     & 692$^{e}$    &  229$^{+3}_{-4}$  \\
V2051 Oph &DN  & 0.0624    & 14.33 &  0.46 &    0.34   &    0$^{c}$ &   9.70     & 92$^{f}$    &   84$^{+18}_{-24}$    \\
V436 Cen  &DN  & 0.0625    & 14.22 &  0.36 &    0.33   & 0.07$^{c}$ &    9.23    & 263$^{g}$    &   97$^{+18}_{-21}$    \\
V347 Pav  &NL  & 0.0626    & 16.20 &  0.68 &    0.54   &    0$^{a}$ &   11.72    & 40-50$^{h}$  &   79$^{+29}_{-46}$    \\
OY Car	  &DN  & 0.0631    & 14.95 &  0.52 &    0.34   &    0$^{c}$ &    9.81    &  100$^{i}$    &  107$^{+25}_{-32}$  \\
UV Per	  &DN  & 0.0649    & 16.47 &  0.74 &    0.30   &    0$^{c}$ &   10.01    & 263$^{b}$    &  196$^{+50}_{-68}$  \\
SX LMi	  &DN  & 0.0672    & 15.71 &  0.15 &    0.17   &    0$^{a}$ &    7.43    & 440$^{j}$    &  453$^{+14}_{-15}$  \\
IR Gem    &DN  & 0.0684    & 15.22 &  0.34 &    0.34   &    0$^{c}$ &    9.19    & 250$^{d}$    &  160$^{+32}_{-42}$  \\
HT Cas    &DN  & 0.0736    & 14.70 &  0.48 &    0.38   & 0.03$^{c}$ &    9.58    & 125$^{k}$    &  104$^{+27}_{-37}$  \\
VW Hyi    &DN  & 0.0743    & 12.52 &  0.48 &    0.34   & 0.01$^{c}$ &    9.28    &  65$^{d}$    &   44$^{+10}_{-14}$   \\
Z Cha     &DN  & 0.0745    & 13.97 &  0.40 &    0.25   &    0$^{c}$ &    8.45    &  130$^{i}$    &  127$^{+22}_{-25}$   \\
WX Hyi    &DN  & 0.0748    & 13.48 &  0.24 &    0.28   &    0$^{c}$ &    8.23    & 100$^{l}$    &  113$^{+17}_{-19}$  \\
BK Lyn    &NL  & 0.0750    & 14.48 &  0.02 &    0.10   &    0$^{a}$ &    6.32    &              &  428$^{+18}_{-17}$  \\
RZ Leo    &DN  & 0.0760    & 16.34 &  0.67 &    0.28   &    0$^{a}$ &    9.30    & 246$^{b}$    &  255$^{+63}_{-84}$  \\
V503 Cyg  &DN  & 0.0777    & 16.37 &  1.08 &    0.09   &    0$^{c}$ &    8.91    &              &  310$^{+69}_{-90}$  \\
DV UMa    &DN  & 0.0859    & 16.89 &  1.03 &    0.07   &    0$^{a}$ &    8.40    & 391$^{b}$    &  499$^{+106}_{-135}$  \\
HU Aqr    &DN  & 0.0868    & 14.18 &  0.28 &    0.26   &    0$^{a}$ &    7.83    & 231$^{b}$    &  186$^{+30}_{-37}$  \\
QS Tel    &NL  & 0.0972    & 14.29 &  0.52 &    0.24   &    0$^{a}$ &    8.06    & 170$^{m}$    &  176$^{+39}_{-51}$  \\
V592 Cas  &NL  & 0.1151    & 12.29 &  0.04 &    0.07   & 0.25$^{n}$ &    4.53    & 360$^{n}$    &  323$^{+19}_{-19}$ \\
V442 Oph  &NL  & 0.1243    & 13.33 &  0.1  &    0.11   & 0.22$^{c}$ &    4.90    &              &  443$^{+5}_{-6}$ \\
AH Men    &NL  & 0.1272    & 12.48 &  0.41 &    0.18   &    0$^{a}$ &    6.60    & 150$^{o}$    &  150$^{+28}_{-34}$ \\
{\bf DN Gem}   & N   & 0.1278    & 15.43 &  0.20 &   -0.03   & 0.20$^{e}$ &    4.08    & 1660$^{e}$   & 1722$^{+115}_{-108}$ \\
MV Lyr    &NL  & 0.1323    & 15.86 &  0.52 &    0.16   &    0$^{c}$ &    6.68    & 322$^{i}$    &  686$^{+140}_{-177}$ \\
BG CMi    &NL  & 0.1347    & 14.55 &  0.27 &    0.14   &    0$^{c}$ &    5.84    &              &  553$^{+75}_{-86}$  \\
SW Sex    &NL  & 0.1349    & 14.21 &  0.22 &    0.10   &    0$^{c}$ &    5.41    & 290$^{p}$    &  575$^{+55}_{-61}$  \\
TT Ari    &NL  & 0.1376    & 11.00 &  0.09 &    0.03   & 0.03$^{c}$ &    4.42    & 185$^{i}$    &  204$^{+1}_{-1}$  \\
WX Ari    &NL  & 0.1394    & 14.41 &  0.31 &    0.18   &    0$^{a}$ &    6.12    & 198$^{b}$    &  454$^{+79}_{-95}$  \\
V1500 Cyg &N   & 0.1396    & 16.12 &  0.63 &    0.12   & 0.43$^{c}$ &    5.56    & 1265$^{q,r,s}$ & 1090$^{+134}_{-152}$  \\
V1315 Aql &NL  & 0.1397    & 14.07 &  0.46 &    0.17   &  0.1$^{c}$ &    6.22    &              &  356$^{+65}_{-80}$  \\
V533 Her  &N   & 0.147     & 14.71 &  0.05 &    0.02   & 0.03$^{c}$ &    4.06    & 1250$^{t}$   & 1332$^{+13}_{-13}$  \\
AO Psc    &NL  & 0.1496    & 13.46 &  0.15 &    0.22   &  0.02$^{u}$ &   5.79    & 250$^{l}$    &  339$^{+56}_{-68}$  \\
AB Dra    &DN  & 0.152     & 13.62 &  0.33 &    0.19   &  0.1$^{c}$ &    5.80    & 400$^{d}$    &  353$^{+61}_{-74}$  \\
BZ Cam    &NL  & 0.1536    & 13.36 &  0.21 &    0.12   & 0.05$^{v}$ &    5.08    & 830$^{w}$    &  443$^{+49}_{-56}$  \\
IP Peg    &DN  & 0.1582    & 12.60 &  0.63 &    0.25   &    0$^{c}$ &    7.19    & 200$^{d}$    &  121$^{+36}_{-52}$  \\
LX Ser    &NL  & 0.1584    & 13.93 &  0.16 &    0.12   &    0$^{c}$ &    4.99    & 210$^{l}$    &  613$^{+70}_{-78}$  \\
VY For    &NL  & 0.1586    & 15.59 &  0.59 &    0.13   &    0$^{a}$ &    6.17    &              &  765$^{+169}_{-216}$  \\
CY Lyr    &DN  & 0.1591    & 13.63 &  0.26 &    0.15   & 0.18$^{c}$ &    5.05    & 330$^{x}$    &  483$^{+59}_{-67}$  \\
CM Del    &DN  & 0.162     & 13.43 &  0.19 &    0.11   & 0.09$^{c}$ &    4.73    &              &  531$^{+52}_{-56}$  \\
KT Per    &DN  & 0.1627    & 13.31 &  0.49 &    0.20   & 0.18$^{c}$ &    5.95    & 245$^{d}$    &  275$^{+58}_{-73}$  \\
AR And    &DN  & 0.163     & 14.59 &  0.59 &    0.27   & 0.02$^{c}$ &    7.08    & 380$^{b}$    &  316$^{+95}_{-135}$  \\
CN Ori    &DN  & 0.1632    & 13.81 &  0.50 &    0.20   &    0$^{c}$ &    6.41    & 295$^{d}$    &  301$^{+74}_{-100}$  \\
X Leo     &DN  & 0.1644    & 14.29 &  0.40 &    0.29   &    0$^{c}$ &    6.77    & 350$^{d}$    &  318$^{+89}_{-123}$  \\
VW Vul    &DN  & 0.1687    & 13.52 &  0.25 &    0.11   & 0.15$^{c}$ &    4.63    & 605$^{d}$    &  566$^{+58}_{-64}$  \\
UZ Ser    &DN  & 0.173     & 14.03 &  0.48 &    0.35   & 0.33$^{d}$ &    6.57    & 280$^{d}$    &  271$^{+75}_{-105}$  \\
GY Cnc    &DN  & 0.1754    & 13.96 &  0.57 &    0.27   &    0$^{a}$ &    6.92    & 320$^{y}$    &  255$^{+79}_{-114}$  \\
WW Cet    &DN  & 0.1758    & 11.08 &  0.14 &    0.11   & 0.03$^{c}$ &    4.53    & 190$^{d}$    &  202$^{+22}_{-24}$  \\
CW Mon    &DN  & 0.1766    & 13.87 &  0.53 &    0.32   & 0.06$^{c}$ &    7.02    & 290$^{i}$    &  229$^{+73}_{-106}$  \\

\hline
\end{tabular}
}
\end{center}
\end{table*}

\begin{table*}
\contcaption{}
\begin{center}
\tiny{
\begin{tabular}{lccccccccc}
\hline
Name      &Type& $P_{orb}(d)$ &  $J$  & $J-H$ & $H-K_{s}$ &  $E(B-V)$  & $M_{Jc}$ &   $d(pc)$    &  $d_{c}(pc)$ \\
   (1)    &(2) &   (3)     &  (4)  &   (5) &    (6)    &    (7)     &     (8)     &      (9)     &    (10)           \\
\hline
TW Vir    &DN  & 0.1827    & 13.35 &  0.27 &    0.14   &    0$^{c}$ &    5.07 & 500$^{d}$     &  452$^{+77}_{-94}$  \\
DQ Her    &N   & 0.1936    & 13.60 &  0.32 &    0.20   & 0.05$^{e}$ &    5.35 & 363$^{e}$     &  437$^{+94}_{-119}$  \\
UX UMa    &NL  & 0.1967    & 12.76 &  0.35 &    0.14   & 0.02$^{c}$ &    5.08 & 340$^{l}$     &  340$^{+67}_{-83}$  \\
V345 Pav  &NL  & 0.1981    & 12.15 &  0.39 &    0.09   &  0$^{a}$   &    4.80 &               &  295$^{+52}_{-62}$   \\
{\bf BT Mon}    &N   & 0.3338    & 14.40 &  0.44 &    0.24   & 0.20$^{e}$ &    4.28 & 1914$^{e}$    &  972$^{+286}_{-406}$   \\
FO Aqr    &NL  & 0.2021    & 12.87 &  0.13 &    0.24   & 0$^{c}$    &    5.16 &               &  349$^{+74}_{-93}$   \\
{\bf T Aur}     &N   & 0.2044    & 16.14 &  0.28 &    0.16   & 0.30$^{e}$ &    4.31 & 991$^{e}$     &  772$^{+108}_{-127}$   \\
{\bf V446 Her}  &N   & 0.207     & 15.39 &  0.59 &    0.11   & 0.25$^{e}$ &    4.80 &  1762$^{e}$   & 1185$^{+224}_{-277}$   \\
RX And    &DN  & 0.2099    & 12.45 &  0.71 &    0.19   & 0.02$^{c}$ &    6.16 & 200$^{d}$     &  180$^{+55}_{-78}$   \\
HR Del	  &N   & 0.2142    & 12.32 &  0.05 &    0.06   & 0.16$^{e}$ &    3.12 & 673$^{e}$     &  648$^{+28}_{-28}$ \\
PQ Gem	  &NL  & 0.2164    & 13.49 &  0.29 &    0.20   &    0$^{a}$ &    5.18 &               &  459$^{+108}_{-142}$   \\
HL CMa    &DN  & 0.2168    & 11.64 &  0.19 &    0.22   &    0$^{c}$ &    4.99 &  80$^{l}$     &  214$^{+47}_{-60}$     \\
AY Psc    &DN  & 0.2173    & 14.52 &  0.50 &    0.02   &    0$^{a}$ &    4.31 &  565$^{b}$    & 1101$^{+179}_{-214}$  \\
EZ Del    &DN  & 0.2234    & 15.08 &  0.30 &    0.08   & 0.16$^{z}$ &    3.82 &  1000$^{z}$   & 1679$^{+212}_{-243}$  \\
V347 Pup  &NL  & 0.2319    & 13.13 &  0.65 &    0.19   & 0.05$^{aa}$ &    5.73 &  470$^{aa}$    &  296$^{+89}_{-126}$  \\
DO Leo    &DN  & 0.2345    & 15.93 &  0.73 &   -0.02   &    0$^{a}$ &    4.45 &  878$^{b}$    & 1971$^{+397}_{-496}$  \\
TX Col    &NL  & 0.2383    & 13.63 &  0.26 &    0.20   & 0.05$^{c}$ &    4.73 &               &  591$^{+135}_{-175}$   \\
AH Eri    &DN  & 0.2391    & 15.92 &  0.58 &    0.40   &    0$^{a}$ &    7.12 & 160$^{b}$     &  576$^{+232}_{-390}$   \\
LL Lyr    &DN  & 0.2491    & 15.42 &  0.53 &    0.25   & 0.06$^{z}$ &    5.63 & 960$^{z}$     &  887$^{+277}_{-402}$   \\
{\bf XY Ari}    &NL  & 0.2527    & 16.14 &  1.88 &    0.90   & 3.7$^{bb}$  &    5.93 & 270$^{bb}$     &  243$^{+82}_{-123}$   \\
TZ Per    &DN  & 0.2629    & 13.09 &  0.58 &    0.11   & 0.27$^{c}$ &    4.17 & 435$^{d}$     &  545$^{+116}_{-147}$    \\
BV Pup    &DN  & 0.265     & 13.34 &  0.45 &    0.10   & 0$^{c}$    &    4.30 & 630$^{d}$     &  642$^{+145}_{-188}$    \\
TT Crt    &DN  & 0.2684    & 13.87 &  0.48 &    0.21   & 0$^{d}$    &    5.16 & 500$^{d}$     &  554$^{+165}_{-235}$    \\
V426 Oph  &DN  & 0.2853    & 11.00 &  0.50 &    0.17   & 0.08$^{c}$ &    4.58 & 202$^{cc}$     &  186$^{+51}_{-69}$      \\
J0813+4528 &CV  & 0.289    & 15.94 &  0.68 &    0.11   & 0.05$^{z}$ &    4.64 & 2100$^{z}$    & 1779$^{+499}_{-695}$      \\
EM Cyg    &DN  & 0.2909    & 11.74 &  0.40 &    0.18   & 0.03$^{c}$ &    4.50 & 350$^{d}$     &  277$^{+75}_{-102}$     \\
AC Cnc    &NL  & 0.3005    & 13.08 &  0.38 &    0.10   & 0$^{c}$    &    3.84 & 400$^{l}$     &  704$^{+161}_{-208}$    \\
RY Ser    &DN  & 0.3009    & 13.70 &  0.63 &    0.18   & 0.39$^{z}$ &    4.21 & 620$^{z}$     &  673$^{+174}_{-235}$    \\
V363 Aur  &NL  & 0.3212    & 13.30 &  0.33 &    0.16   & 0.13$^{c}$ &    3.69 & 900$^{dd}$     &  790$^{+187}_{-245}$    \\
V1309 Ori &NL  & 0.3326    & 14.22 &  0.56 &    0.15   &    0$^{a}$ &    4.40 & 550$^{ee}$    &  920$^{+279}_{-399}$    \\
CH UMa    &DN  & 0.3432    & 12.71 &  0.50 &    0.17   & 0.06$^{z}$ &    4.17 & 480$^{z}$     &  498$^{+146}_{-208}$    \\
SY Cnc    &DN  & 0.38      & 11.27 &  0.22 &    0.13   & 0$^{c}$    &    3.04 & 300$^{l}$     &  443$^{+104}_{-137}$    \\
Q Cyg     &N   & 0.4202    & 13.54 &  0.29 &    0.14   & 0.44$^{c}$ &    2.12 & 2188$^{e}$    & 1610$^{+306}_{-376}$    \\
\hline
\end{tabular}\\
{
a: assumed value (see text), b: \citet{Sproatsetal1996}, c: \citet{BruchandEngel1994}, 
d: \citet{UrbanandSion2006}, e: \citet{Selvelli2004}, 
f: \citet{SaitoandBaptista2006}
g: \citet{NadalinandSion2001}
h: \citet{Ramsay2004}, 
i: \citet{Verbuntetal1997}
j: \citet{Wagneretal1998}
k: \citet{Woodetal1992}
l: \citet{Patterson1984}, 
m: \citet{Gerkeetal2006}, 
n: \citet{Tayloretal1998}, 
o: \citet{Gaensicke1999}, 
p: \citet{Grootetal2001}, 
q: \citet{Esenoglu1997}, 
r: \citet{Cohen1985}, 
s: \citet{Duerbeck1981}, 
t: \citet{Gill2000}, 
u: \citet{HellierandvanZyl2005}, 
v: \citet{Prinjaetal2000}, 
w: \citet{RingwaldandNaylor1998}, 
x: \citet{ThorstensenandTaylor1998}, 
y: \citet{Gaensickeetal2000}, 
z: \citet{Thorstensenetal2004}, 
aa: \citet{Thoroughgoodetal2005}, 
bb: \citet{Littlefairetal2001}, 
cc: \citet{Hessman1988}, 
dd: \citet{Warner1995}, 
ee: \citet{Staudeetal2001}
}
}
\end{center}
\end{table*}

\section{Discussion}

We have suggested an absolute magnitude calibration for CVs based on the trigonometric parallaxes 
and {\em 2MASS} observations. The calibration equation covers a wide range of periods 
and colours. The mean error in the absolute magnitude $M_{J}$ is 0.22. However, there is a considerable 
scatter for some systems. 

Large deviations from the PLCs relation are seen mostly in faint systems. 
The source of the deviations can not be the colour excess $E(B-V)$, since the deviated systems 
are located in the middle and higher Galactic latitudes ($\mid b \mid$ $\geq$ $27^{o}$) and they 
are relatively close objects. 
{\bf Also, since they are relatively close objects, it is unlikely that 
they are due to parallax errors. It is therefore more likely that these deviations come from the 
intrinsic properties of the systems. Although for the systems GW Lib and EF Eri the magnitude flags 
in {\em 2MASS} are (ABC) and (DBC), respectively, which indicates low-quality observations, 
the fact that the largest deviations come from systems with the faintest absolute magnitudes suggests 
that here the disc makes a more dominant contribution relative to the donor star than in systems 
with brighter donor stars. Indeed,} the nova-like system EF Eri has possibly a substellar 
secondary (\citealt{Beuermannetal2000}) and the deviation of this system from the PLCs relation can 
then be attributed to its very faint substellar component. Other deviated systems (GW Lib, T Leo, 
VY Aqr and Z Cam) are all dwarf novae. {\bf Disc or donor activity} or a third component of the binary 
system can affect the magnitude of the system. We used AAVSO's Light Curve 
Generator\footnote{http://www.aavso.org/data/lcg/} to find the activity stage of these systems 
during the {\em 2MASS} observations. Unfortunately, only Z Cam has enough visual observations to 
conclude that its activity stage. We found that this system was at the beginning of the decline 
from an outburst during the {\em 2MASS} observations. 

As for the non or {\bf less-deviating} dwarf novae from the PLCs relation, only SU UMa was found in 
a superoutburst maximum from the AAVSO's Light Curve Generator. Other dwarf novae were either in 
quiescence (WZ Sge, EX Hya, V893 Sco, YZ Cnc, U Gem, SS Aur, RU Peg) or in the decline branch 
from an outburst (AH Her and SS Cyg) like Z Cam during the {\em 2MASS} observations. These systems are 
located in or very near the 99$\%$ confidence limit without regarding their activity stage. So, it is 
possible to say that activity stage does not affect the location of a CV in the PLCs relation. However, 
{\bf it seems} that finding considerable deviations from the PLCs relation mostly for some dwarf novae 
{\bf is not} a coincidence {\bf (Interestingly note that the nova-like star GP Com, in which a CO white dwarf 
is acreeting from a helium degenerate (\citealt{Morales-Ruedaetal2003}), obeys the relation very well).}

We compared the distances obtained from the PLCs relation with those found by various other methods in 
Figure 2. Although Figure 2 shows that the distances found from the PLCs relation are generally somewhat
{\bf smaller} than those found by other methods, {\bf for the nova-like systems} the PLCs relation yields distances 
longer than other methods. The distance inferred from the trigonometric parallax of HR Del is very different 
than found from its shell parallax (760 pc, \citealt{Duerbeck1999}). This is why we did not include this 
system in our data sample listed in Table 1. Its distance found from the PLCs relation is, {bf however}, much closer 
to that obtained from the shell parallax. 

{\bf In view of the above results, we} suggest that the PLCs relation can be a useful statistical tool to calculate 
the distances of CVs from their {\em 2MASS} observations since the PLCs relation 
{\bf has been calibrated} with the most reliable distance estimation method (trigonometric parallax). Distances 
calculated from the PLCs relation can give clues for astrometric observations of these systems, as well. 
Finally, it should be stated that future astrometric observations of CVs such as {\em GAIA} and 
{\em SIM} missions, will refine the PLCs relation.

\section{Acknowledgments}
{\bf We thank the anonymous referee for a thorough report and useful comments that helped improving an early 
version of the paper.} We acknowledge the observers of the AAVSO who made the observations that were 
used to check activity stages during the {\em 2MASS} observations of the dwarf novae in this study. 
This publication makes use of data products from the Two Micron All Sky Survey, which is a joint 
project of the University of Massachusetts and the Infrared Processing and Analysis Center/California 
Institute of Technology, funded by the National Aeronautics and Space Administration and the National 
Science Foundation. Part of this work was supported by the Research Fund of the University of Istanbul, 
Project Numbers: BYP-724/24062005 and BYP-738/07072005.

\end{document}